\documentclass[prb,onecolumn,12pt,superscriptaddress]{revtex4-1}
\usepackage{epsfig}
\usepackage{amssymb}
\usepackage{threeparttable}
\usepackage{graphicx}
\usepackage{natbib}
\usepackage{longtable}
\usepackage{txfonts}
\usepackage{color}

\begin{document}
\title{Stability and Strength of Atomically Thin Borophene from First Principles Calculations}
\author{Bo Peng}
\affiliation{Shanghai Ultra-precision Optical Manufacturing Engineering Center, Department of Optical Science and Engineering, Fudan University, Shanghai 200433, China}
\author{Hao Zhang}
\email{zhangh@fudan.edu.cn}
\affiliation{Shanghai Ultra-precision Optical Manufacturing Engineering Center, Department of Optical Science and Engineering, Fudan University, Shanghai 200433, China}
\author{Hezhu Shao}
\affiliation{Ningbo Institute of Materials Technology and Engineering, Chinese Academy of Sciences, Ningbo 315201, China}
\author{Zeyu Ning}
\affiliation{Shanghai Ultra-precision Optical Manufacturing Engineering Center, Department of Optical Science and Engineering, Fudan University, Shanghai 200433, China}
\author{Yuanfeng Xu}
\affiliation{Shanghai Ultra-precision Optical Manufacturing Engineering Center, Department of Optical Science and Engineering, Fudan University, Shanghai 200433, China}
\author{Hongliang Lu}
\affiliation{State Key Laboratory of ASIC and System, Institute of Advanced Nanodevices,School of Microelectronics, Fudan University, Shanghai 200433, China.}
\author{David Wei Zhang}
\affiliation{State Key Laboratory of ASIC and System, Institute of Advanced Nanodevices,School of Microelectronics, Fudan University, Shanghai 200433, China.}
\author{Heyuan Zhu}
\affiliation{Shanghai Ultra-precision Optical Manufacturing Engineering Center, Department of Optical Science and Engineering, Fudan University, Shanghai 200433, China}

\begin{abstract}
A new two-dimensional (2D) material, borophene (2D boron sheet), has been grown successfully recently on single crystal Ag substrates by two parallel experiments [Mannix \textit{et al., Science}, 2015, \textbf{350}, 1513] [Feng \textit{et al., Nature Chemistry}, 2016, \textbf{advance online publication}]. Three main structures have been proposed ($\beta_{12}$, $\chi_3$ and striped borophene). However, the stability of three structures is still in debate. Using first principles calculations, we examine the dynamical, thermodynamical and mechanical stability of $\beta_{12}$, $\chi_3$ and striped borophene. Free-standing $\beta_{12}$ and $\chi_3$ borophene is dynamically, thermodynamically, and mechanically stable, while striped borophene is dynamically and thermodynamically unstable due to high stiffness along $a$ direction. The origin of high stiffness and high instability in striped borophene along $a$ direction can both be attributed to strong directional bonding. This work provides a benchmark for examining the relative stability of different structures of borophene.
\end{abstract}

\maketitle

\section{Introduction}

Recent years have witnessed many breakthroughs in research on two-dimensional (2D) materials due to their potential applications in next-generation electronic and energy conversion devices \cite{Zhang2005,Balandin2008,Geim2009,Liu2011a,Novoselov2012,Xu2013,Xu2014a,Issi2014,Luican-Mayer2014,Davila2014,Ferrari2015,Peng2016b,Peng2016a,Peng2016c}. Recently, a new type of 2D material, borophene (2D boron sheet) \cite{Piazza2014}, has been successfully grown on single crystal Ag(111) substrates by two parallel experiments \cite{Mannix18122015,Feng2016}. Although various proposals of stable 2D boron sheets and quasiplanar boron clusters have been made \cite{Boustani1997,Zhai2003,Zhai2003a,Tang2007,Lau2007,Yang2008a,Penev2012,Liu2013a,Liu2013b,Zhou2014,Zhai2014,Li2015,Yuan2015,Zhang2015c,Wang2016d,Zhou2016,Carrete2016}, three main structures ($\beta_{12}$, $\chi_3$ and striped borophene) have been observed by scanning tunneling microscopy in these two experiments: $\beta_{12}$ and $\chi_3$ borophene has planar structure with periodic holes \cite{Feng2016}, while striped borophene has buckled structure with anisotropic corrugation \cite{Mannix18122015}. The following first principles calculations have predicted that striped borophene possesses remarkable mechanical properties \cite{Wang2016,Pang2016}, which may rival graphene \cite{Mannix18122015}. However, phonon instability in striped borophene is observed \cite{Peng2016d}, which may challenge previous results demonstrating that borophene is stiffer than graphene along $a$ direction \cite{Mannix18122015,Wang2016,Pang2016}.

Theoretical investigation of the formation of boron sheet on Ag(111) surface has demonstrated that stable boron sheet should contain 1/6 vacancies in a striped pattern \cite{Xu2016}. This is consistent with previous theoretical studies indicating that planar boron sheets with vacancies are more stable \cite{Tang2007,Yang2008a,Penev2012,Yu2012,Lu2013,Liu2013a,Li2015,Yuan2015,Zhang2015c}. The ground state of 2D boron is in debate \cite{Tang2007,Yu2012,Wu2012,Lu2013,Zhang2015c,Mannix18122015,Meng2016,Feng2016,Wang2016,Pang2016,Feng2015,Xu2016,Gao2016,Liu2016a}. These debates raise several questions: (i) Are these three structures stable? (ii) What is the relative stability of $\beta_{12}$, $\chi_3$ and striped borophene? (iii) Could these structures possess high hardness in thermodynamic aspect? In fact, due to the structural complexity of boron, even the relative stability of $\alpha-$ and $\beta-$rhombohedral boron has been discussed for over 30 years \cite{Bullett1982,Jemmis2001,Jemmis2001a,Imai2002,Prasad2005,Ogitsu2013,Solozhenko2013}. Thus a systematic investigation is needed to examine the stability and strength of these three structures.

When discussing the stability of crystal structures, it is important to distinguish dynamical, thermodynamical and mechanical stability \cite{Zhang2012,Zhou2014a}. A material is dynamically stable when no imaginary phonon frequencies exist. Thermodynamical stability can be described by the Helmholtz free energy \cite{Setten2007}, which demonstrates how phonons determine the relative stability at finite temperatures. Regarding mechanical stability, the Born-Huang criteria for elastic constants must be fullfilled \cite{Born1954,Wu2007}. The Born-Huang mechanical stability criteria provide a \textit{necessary} condition for the dynamical stability, but not a \textit{sufficient} one \cite{Zhou2014a}. Therefore, although the mechanical properties of striped borophene have been studied intensively \cite{Mannix18122015,Wang2016,Pang2016}, the questions of stability still remain unanswered. In this work, the dynamical, thermodynamical and mechanical stabilities of $\beta_{12}$, $\chi_3$ and striped borophene are evaluated using first principles calculations. We further study for the first time the mechanical properties of $\beta_{12}$ and $\chi_3$ borophene, and compare the stiffness of three distinguished structures. The bonding characteristics are also examined to understand the stability and strength of striped borophene.

\section{Results and discussion}

\begin{figure*}
\centering
\includegraphics[width=0.9\linewidth]{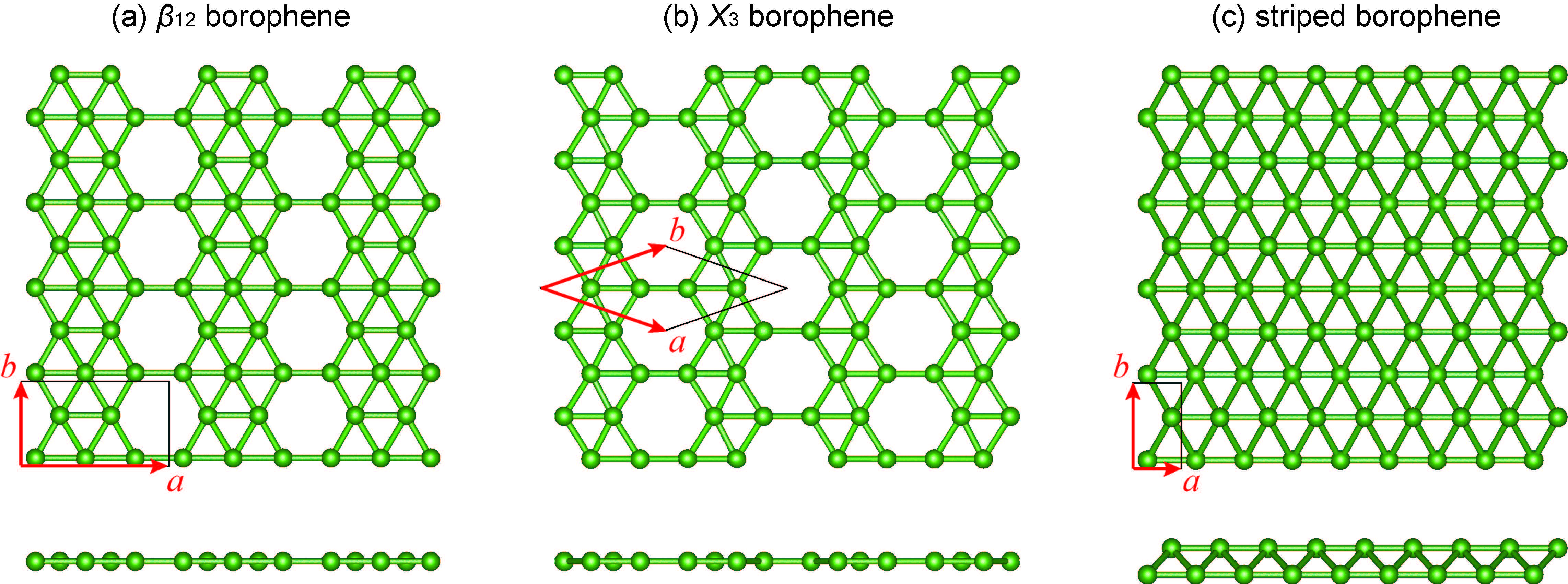}
\caption{Top view and side view of (a) $\beta_{12}$, (b) $\chi_3$ and (c) striped borophene.}
\label{structure}
\end{figure*}

\begin{figure*}
\centering
\includegraphics[width=0.98\linewidth]{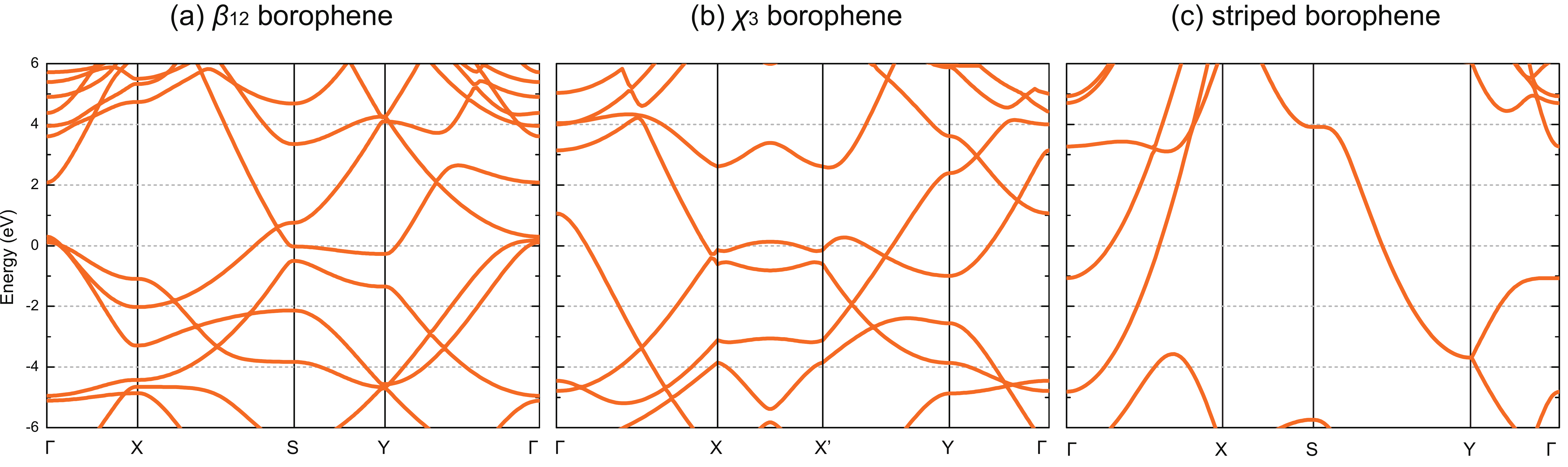}
\caption{Calculated band structures of (a) $\beta_{12}$ borophene and (b) striped borophene along different symmetry lines.}
\label{band} 
\end{figure*}

\subsection{Crystal structure and dynamical stability}

Three structures of borophene are shown in Fig.~\ref{structure}. For $\beta_{12}$ ($\chi_3$) borophene, five (four) B atoms in the unit cell are arranged in the same plane. For striped borophene, there is no corrugation along $a$ direction, while a vertical buckling along the $b$ direction is observed. The major difference between striped borophene and the other two structures is the absence of vacancies. The introduction of vacancies leads to lower cohesive energy, as shown in Table~\ref{lattice}, which is in good agreement with previous results \cite{Tang2007,Yu2012}. The optimized geometries of three structures are listed in Table~\ref{lattice}. The predicted $a$ of striped borophene corresponds to 1/3 the $a$ observed in the experiment, while the lattice constants of $\beta_{12}$ and $\chi_3$ borophene is in excellent agreement with experimental results \cite{Mannix18122015,Feng2015}. Thus the reliability of the present calculations is confirmed. Fig.~\ref{band} presents the calculated band structures of all three phases, which agree well with previous theoretical results \cite{Penev2016}. It should be noticed that due to highly anisotropic crystal structure, striped borophene shows anisotropic metallic behaviour. 

\begin{table*}
\centering
\caption{Calculated lattice constant $a$ and $b$, buckling height $h$, cohesive energy $E_c$, and zero point energy ZPE of three structures of borophene. Previous experimental data are also listed for comparison.}
\begin{tabular}{ccccccc}
\hline
 & Space group & $a$ & $b$ & $h$ & $E_c$ & ZPE \\
 &  & (\AA) & (\AA) & (\AA) & (eV/atom) & (eV/atom) \\
\hline
$\beta_{12}$ borophene & $Pmm2$ & 5.07 & 2.93 & 0 & -6.147 & 0.116 \\
 & - & 5.0 \cite{Feng2016} & 2.9 \cite{Feng2016} & - & - \\
$\chi_3$ borophene & $Cmmm$ & 4.45 & 4.45 & 0 & -6.159 & 0.114 \\
 & - & 4.3 \cite{Feng2016} & 4.3 \cite{Feng2016} & - & - \\
striped borophene & $Pmmn$ & 1.613 & 2.864 & 0.911 & -6.099 & 0.109 \\
 & - & 5.1$\pm$0.2 \cite{Mannix18122015} & 2.9$\pm$0.2 \cite{Mannix18122015} & - & - \\
\hline
\end{tabular}
\label{lattice}
\end{table*}

\begin{figure*}
\centering
\includegraphics[width=\linewidth]{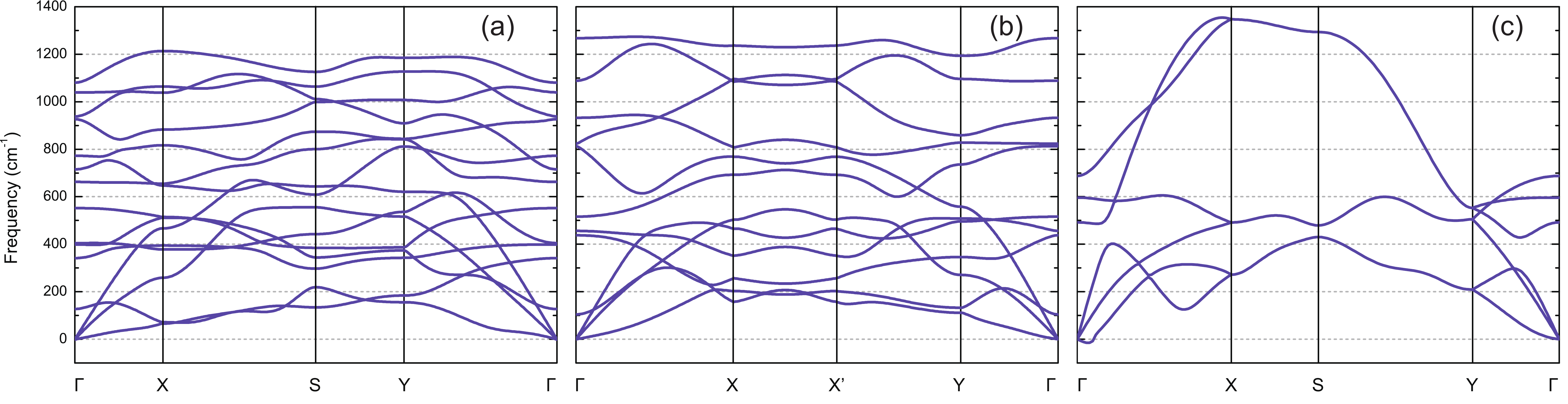}
\caption{Phonon dispersion for (a) $\beta_{12}$, (b) $\chi_3$ and (c) striped borophene along different symmetry lines.}
\label{phonon} 
\end{figure*}

For newly proposed 2D materials, stability is an important aspect for experimental realization and large-scale production. We investigate the dynamical stability of these three structures in Fig.~\ref{phonon}. No imaginary vibrating mode is seen for $\beta_{12}$ and $\chi_3$ borophene, which demonstrates that these structures are kinetically stable at 0 K. However, for striped borophene, the vibrational frequencies become imaginary in the long-wavelength limit along $\Gamma$-X direction, showing its dynamical instability for long-wavelength acoustic vibrations \cite{Mannix18122015,Peng2016d}. The imaginary frequencies remain even when employing a larger supercell with a higher convergence criterion (especially the \textbf{k}-mesh). In fact, a recent study has found that free-standing striped borophene is dynamically instable even under high tensile stress \cite{Wang2016}.

\subsection{Thermodynamical stability}

For light elements such as boron, phonons play an important role in determining the thermodynamical stability of crystals both at 0 K and at finite temperatures \cite{Setten2007}. Using phonon frequencies in the whole Brillouin zone, we further examine the thermodynamical stability of three structures of borophene by calculating the Helmholtz free energy $F$ \cite{Setten2007},
\begin{equation}
F = E_{tot} + \frac{1}{2} \sum\limits_{\textbf{q}j}\hbar\omega_{\textbf{q}j}+k_BT\sum\limits_{\textbf{q}j}\ln[1-\exp(-\hbar \omega_{\textbf{q}j}/k_BT)],
\end{equation}
where $E_{tot}$ is the total energy of the crystal, and the summation term is the Helmholtz free energy for phonons \cite{Togo2008,Togo2015}. The first summation term is a temperature-free term corresponding to the zero point energy (ZPE) of phonons; and the second summation term is a temperature-dependent term referring to the thermally induced occupation of the phonon modes. The calculated ZPE of $\beta_{12}$, $\chi_3$ and striped borophene are listed in Table~\ref{lattice}, which are the Helmholtz free energies of phonons at 0 K. The inclusion of the ZPE brings the $F$ of $\beta_{12}$, $\chi_3$ and striped borophene to -6.145 eV/atom, -6.159 eV/atom and -6.106 eV/atom, respectively.

\begin{figure}
\centering
\includegraphics[width=0.5\linewidth]{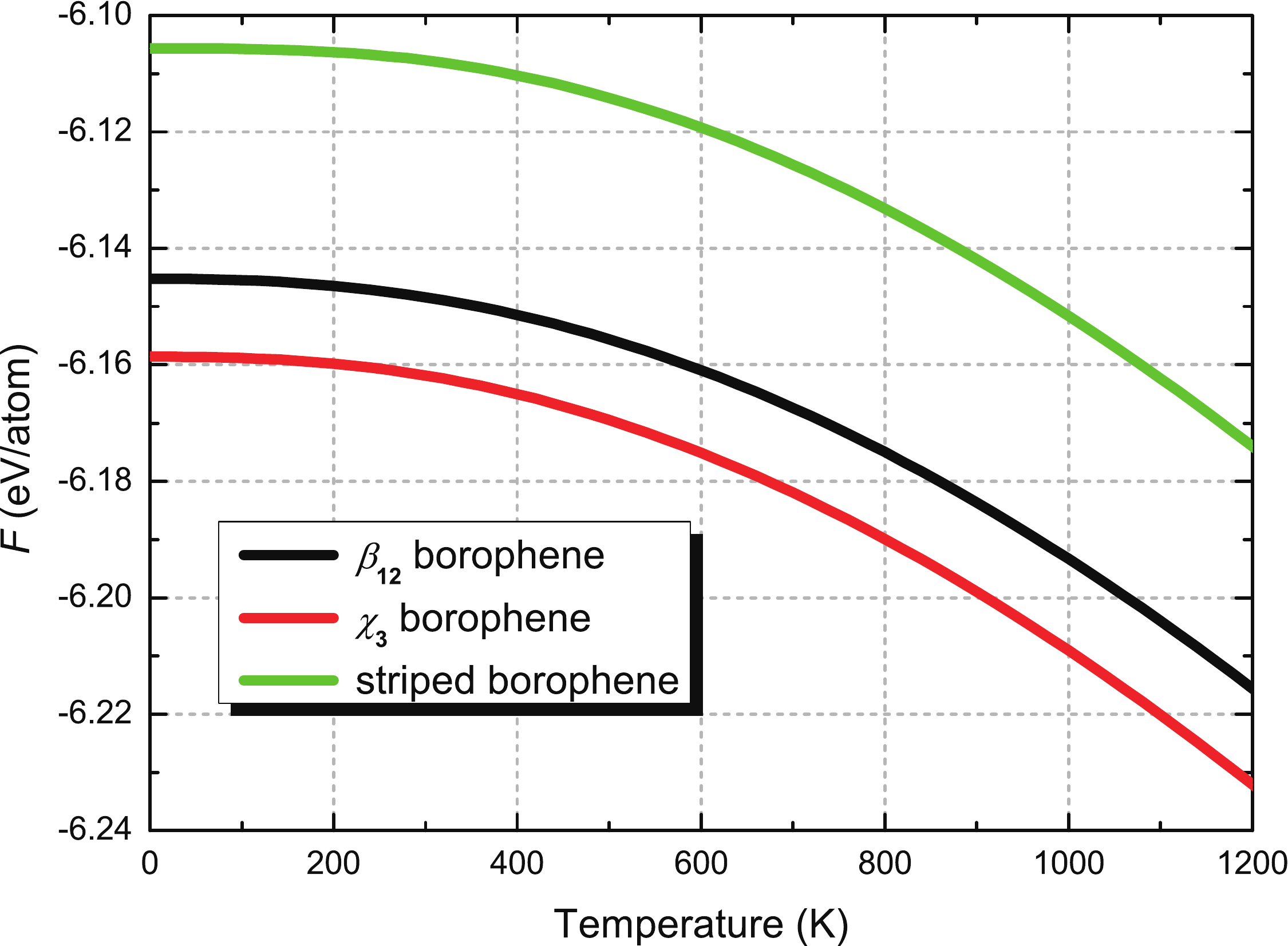}
\caption{Helmholtz free energy as a function of temperature for (a) $\beta_{12}$, (b) $\chi_3$ and (c) striped borophene.}
\label{free} 
\end{figure}

Temperature is also an important thermodynamic variable for determining the stability of materials. At higher temperature, the phonon modes are occupied according to Bose-Einstein statistics. The Helmholtz free energies $F$ as a function of temperature are shown in Fig.~\ref{free}. In the temperature range of 0-1000 K, the Helmholtz free energy of $\chi_3$ borophene is lowest, while the $F$ of $\beta_{12}$ borophene is much lower than that of striped borophene, indicating that $\beta_{12}$ and $\chi_3$ borophene is more thermodynamically stable than striped borophene over a wide temperature range.

In comparison to striped borophene, the thermodynamical stability of $\beta_{12}$ and $\chi_3$ borophene is due to the softness of the phonon modes: The phonon frequencies of $\beta_{12}$ and $\chi_3$ borophene are relatively low, leading to an increase in entropy at high temperatures, thus the structures are more stabilized than that of striped borophene, which is similar to $\alpha-$ and $\beta-$boron \cite{Masago2006}, and $\alpha-$ and $\beta-$tin \cite{Pavone1998}.

\subsection{Mechanical stability}

To investigate the mechanical stability of both structures, we calculate the elastic constants. Due to 3D periodic boundary conditions, the 2D coefficients $C_{ij}^{2D}$ need to be renormalized by the vacuum space between the 2D layers \cite{Blonsky2015}, $i.e.$ $C_{ij}^{2D}=c\ C_{ij}^{3D}$. The calculated elastic constants of both $\beta_{12}$, $\chi_3$ and striped borophene in Table~\ref{stiff} satisfy the corresponding Born stability criteria according to Born-Huang's lattice dynamical theory \cite{Born1954,Wu2007}, indicating both structures are mechanically stable.

\begin{table*}
\centering
\caption{Calculated elastic coefficients $C_{ij}^{2D}$ for three structures of borophene, as well as 2D Young's modulus $E^{2D}$, Poisson's ratio $\nu^{2D}$ parallel and perpendicular to $a$ direction.}
\begin{tabular}{ccccccccc}
\hline
 & $C_{11}^{2D}$ & $C_{12}^{2D}$ & $C_{22}^{2D}$ & $C_{66}^{2D}$ & $E_{a//}^{2D}$ & $E_{a\perp}^{2D}$ & $\nu_{a//}^{2D}$ & $\nu_{a\perp}^{2D}$ \\
& (GPa$\cdot$nm) & (GPa$\cdot$nm) & (GPa$\cdot$nm) & (GPa$\cdot$nm) & (GPa$\cdot$nm) & (GPa$\cdot$nm) \\
\hline
$\beta_{12}$ borophene & 188.1 & 36.0 & 214.3 & 63.5 & 182.0 & 207.5 & 0.17 & 0.19 \\
$\chi_3$ borophene & 194.8 & 36.2 & 187.6 & 70.7 & 187.8 & 180.8 & 0.19 & 0.19 \\
striped borophene & 382.5 & -5.8 & 154.2 & 76.4 & 382.3 & 154.1 & -0.04 & -0.02 \\
 & - & - & - & - & 398 \cite{Mannix18122015} & 170 \cite{Mannix18122015} & -0.04 \cite{Mannix18122015} & -0.02 \cite{Mannix18122015} \\
 & - &-  & - & - & 389 \cite{Wang2016} & 166 \cite{Wang2016} & - & - \\
graphene \cite{Andrew2012} & 352.7 & 60.9 & 352.7 & 145.9 & 342.2 & 342.2 & 0.173 & 0.173 \\
BN \cite{Andrew2012} & 289.8 & 63.7 & 289.8 & 113.1 & 275.8 & 275.8 & 0.220 & 0.220 \\
silicene \cite{Andrew2012} & 68.3 & 23.3 & 68.3 & 22.5 & 60.6 & 60.6 & 0.341 & 0.341 \\
\hline
\end{tabular}
\label{stiff}
\end{table*}

Using the elastic tensor, the mechanical properties such as 2D Young's modulus $E^{2D}$ and the corresponding Poisson's ratio $v^{2D}$ can be calculated \cite{Andrew2012}, as shown in Table~\ref{stiff}. The calculated $E^{2D}$ of striped borophene are in good agreement with other theoretical data \cite{Mannix18122015,Wang2016}. 2D Young's modulus is defined as the ratio between stress and strain, and provides a measure of in-plane stiffness of the solid materials. The mechanical properties of $\beta_{12}$ and $\chi_3$ borophene are similar due to similar in-plane bonding and atomic mass density. As shown in Table~\ref{stiff}, the stiffness of $\beta_{12}$ and $\chi_3$ borophene is lower than that of graphene and monolayer BN, but higher than that of silicene \cite{Andrew2012}.

Negative Possion's ratio is observed in striped borophene, which is due to highly buckled structure \cite{Wang2016}. In addition, the stiffness for striped borophene along $a$ direction is much higher than that along $b$ direction, and even rivals graphene (342.2 GPa$\cdot$nm) \cite{Andrew2012}. In fact, although striped borophene is much stiffer than $\beta_{12}$ and $\chi_3$ borophene, thermodynamically, high stiffness means unstability due to increasing Helmholtz free energy via the increase in vibration frequency \cite{Masago2006}. Both the high stiffness and high instability of striped borophene along $a$ direction can be attributed to strong directional bonding.

\subsection{Bonding characteristics}

\begin{figure*}
\centering
\includegraphics[width=0.8\linewidth]{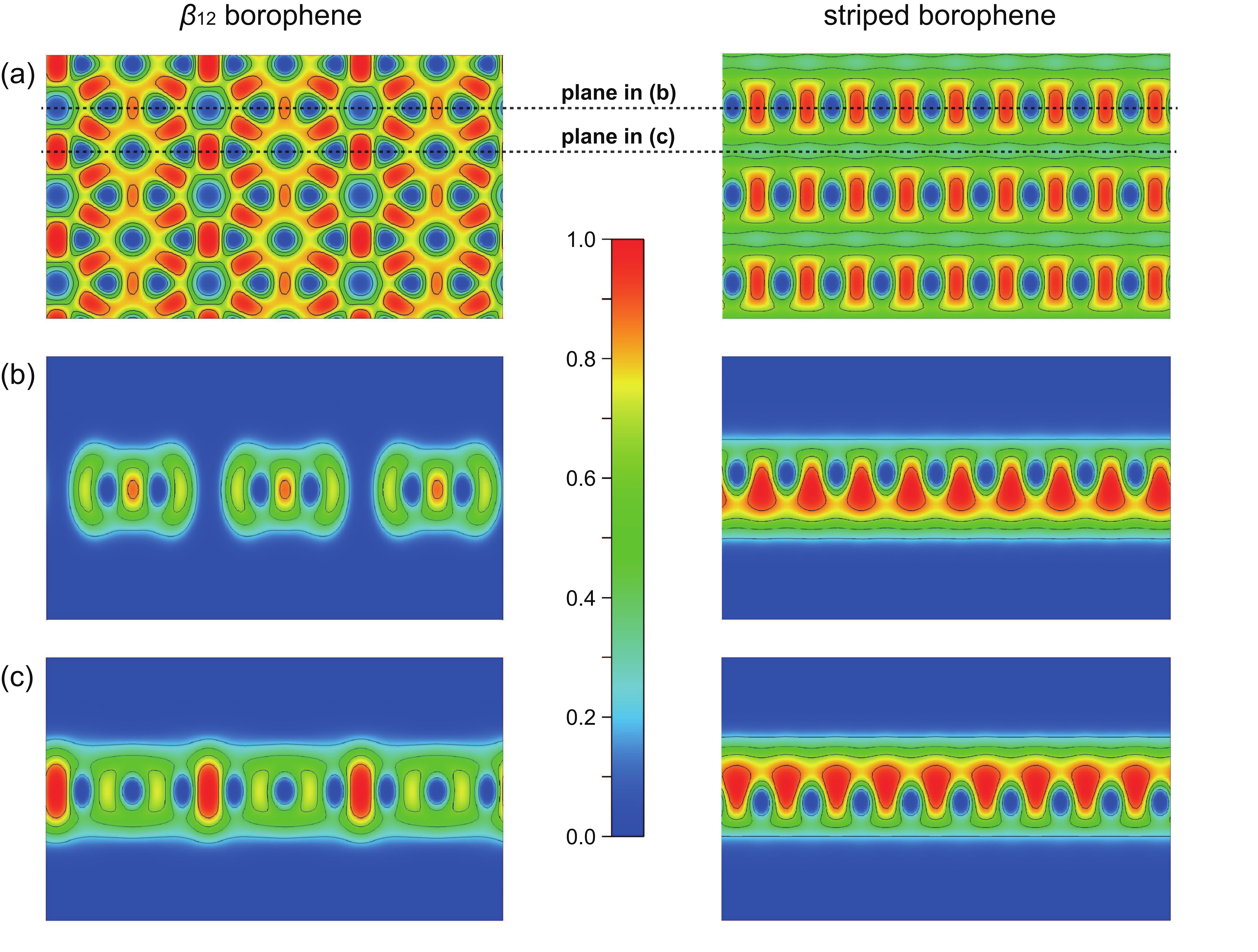}
\caption{(a) Top view of 2D ELF profiles of $\beta_{12}$ and striped borophene, as well as side view in the [010] plane with distance of (b) 1.75 \AA\ (1.72 \AA) and (c) 3.22 \AA\ (3.15 \AA) from origin for $\beta_{12}$ (striped) borophene.}
\label{elf} 
\end{figure*}

To understand the chemical bonding of striped borophene, we calculate its electron localization function (ELF) \cite{Becke1990,Savin1992,Gatti2005,Chen2013} in comparison with $\beta_{12}$ borophene, which has similar 2D orthorhombic structure. A higher value of ELF corresponds to higher electron localization. As shown in Fig.~\ref{elf}, ELF profiles of striped borophene is more anisotropic than that of $\beta_{12}$ borophene. Strong directional bonding may prevent dislocations from forming to accommodate strains and thereby cause the material to be brittle \cite{Fu1990,Ravindrana1997}. This is consistent with previous theoretical studies showing that striped borophene is dynamically unstable even under high tensile stress \cite{Wang2016,Pang2016}. As for $\beta_{12}$ borophene, strong B-B bonds along different directions in the 2D plane of $\beta_{12}$ borophene stabilize the structure. Thus $\beta_{12}$ borophene is more stable than striped borophene considering the bonding characteristics.

\begin{figure}
\centering
\includegraphics[width=0.4\linewidth]{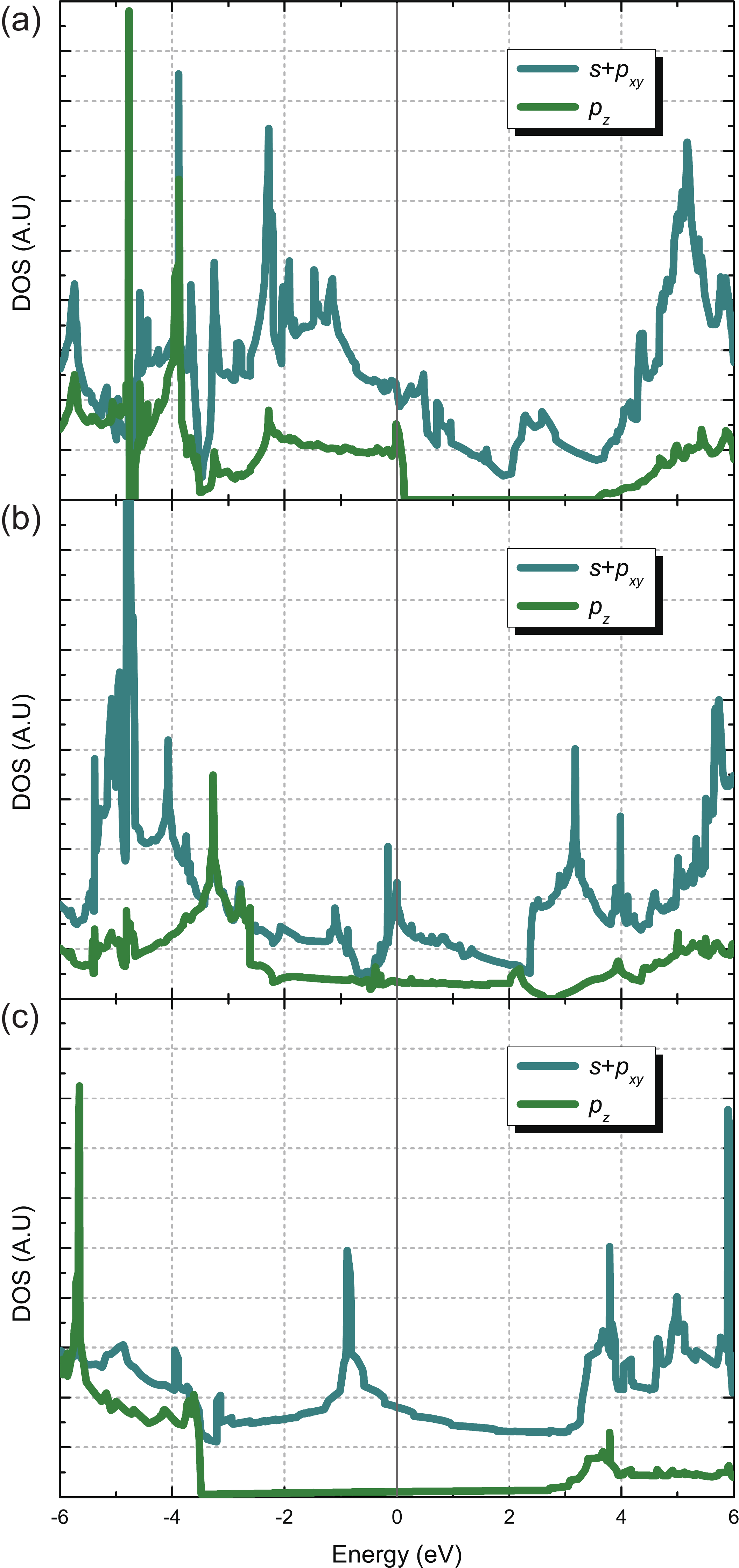}
\caption{Projected DOS for (a) $\beta_{12}$, (b) $\chi_3$ and (c) striped borophene.}
\label{DOS} 
\end{figure}

We further consider the nature of the electronic bonding by calculating the projected density of states (DOS) for three structures of borophene with separated in-plane ($s$, $p_x$ and $p_y$) and out-of-plane ($p_z$) projections in Fig.~\ref{DOS}. Generally, in-plane $sp^2$ bonds ($\sigma$ bonds) are stronger than $sp^3$ bonds ($\pi$ bonds) derived from $p_z$ orbitals. For striped borophene, the $p_z$ projected DOS vanishes from -3.5 to 2.5 eV. As a result, localized $\sigma$ bonds along $a$ direction are observed in ELF profiles. However, these strong $\sigma$ bonds will be destabilized by any flattening of the boron sheet, leading to highly metastable structure \cite{Kunstmann2006}. Thus striped borophene becomes instable under tension along $a$ direction \cite{Wang2016}. In addition, some of the strong in-plane $sp^2$ bonding states are unoccupied in Fig.~\ref{DOS}(c), and subsequently striped borophene tends to accepting electrons to increase its stability \cite{Tang2007}.

\section{Conclusion}

In conclusion, first-principles calculations are performed on 2D borophene sheet to evaluate the dynamical, thermodynamical and mechanical stability of two distinguished structures. Our results show that the free-standing $\beta_{12}$ and $\chi_3$ borophene is thermodynamically, mechanically and dynamically stable, while striped borophene is thermodynamically and dynamically unstable due to high stiffness along $a$ direction. Our calculated ELF shows that the bonding characteristic of striped borophene leads to high stiffness and high instability at the same time.

\section*{Acknowledgement}

This work is supported by the National Natural Science Foundation of China under Grants No. 11374063 and 11404348, and the National Basic Research Program of China (973 Program) under Grant No. 2013CBA01505.

\newpage
\section*{Reference}

%

\end{document}